# SAR Image Despeckling Based on Convolutional Denoising Autoencoder

Zhang Qianqian[1]
PhD student,
College of Information and Electrical Engineering,
China Agricultural University
Beijing, P.R. China
e-mail: qqzhang@cau.edu.cn

Sun Ruizhi[2,*]
Professor,
College of Information and Electrical Engineering,
China Agricultural University
Beijing, P.R. China
*Corresponding author: e-mail: sunruizhi@cau.edu.cn

*Abstract*—In Synthetic Aperture Radar (SAR) imaging, despeckling is very important for image analysis, whereas speckle is known as a kind of multiplicative noise caused by the coherent imaging system. During the past three decades, various algorithms have been proposed to denoise the SAR image. Generally, the BM3D is considered as the state of art technique to despeckle the speckle noise with excellent performance. More recently, deep learning make a success in image denoising and achieved a improvement over conventional method where large train dataset is required. Unlike most of the images SAR image despeckling approach, the proposed approach learns the speckle from corrupted images directly. In this paper, the limited scale of dataset make a efficient exploration by using convolutioal denoising autoencoder (C-DAE) to reconstruct the speckle-free SAR images. Batch normalization strategy is integrated with C-DAE to speed up the train time. Moreover, we compute image quality in standard metrics, PSNR and SSIM. It is revealed that our approach perform well than some others.

*Keywords—SAR image; Image despeckling; Convolutioal denoising autoencoder; Batch normalization* Introduction

Remote-sensing (RS) products play an very important role in current earth observation research procedures where vehicle conveying tremendous information of Earth's surface. Land use management, environment monitoring, agricultural production, to be given several major applications, require the analysis of a very large number of images, far exceeding the processing capacity of general machines or trained human personnel. Most image processing and analysis tasks assume great raw image quality, whereas in practice images are subject to various types and level of noise, which are supposed to provide a reliable access to the information. In (single-look) synthetic aperture radar (SAR) images, a similar noise problem is encountered when coherent speckle, a kind of multiplicative noise, is generated during SAR imaging. SAR is such a mature technology, providing broad-area image at high resolution even in poor weather condition day and night, and supplementing the features sensed by optical images. Therefore, there is an increasing intense quest for reliable despeckling techniques that can succeed in the removal of most speckles without impairing potentially valuable images.

Firstly, we clarify the meaning of SAR denoising in the context same as SAR despeckling. SAR images are analyzed by trained human experts or special programs to extract certain information of interest for the end users without miss radar signal. To this end, SAR image noise is limited to speckle. Indeed, image despeckling has been an active field of research for almost 30 years, and many new algorithms are proposed each year that appear to provide better performance. The availability of noise in the SAR images has obvious influence that complicates the obtaining and analysis process timely[1], as results depend strongly on the implementation of subsequent tasks (object detection, classification, segmentation, and so on). Also, the denoising process is disturbing the quality of the original images which may lead to poor decisions either by humans or machines. Therefore, the goal of noise removal or reduction in the SAR image should take accuracy into high consideration as much as possible.

SAR image despeckling, conventional problem in the field of computer vision, has been extensively studied by researchers. Image filtering techniques are used to remove speckles [2] but the filtering is always combined with multilook integration techniques. Classical spatial filters such as the Lee [3], Frost [4], Kuan [5] and Gamma-MAP [6] filters, can remove speckles to some degree, but they also lead to blurring effects and defects in detail preservation. Recently, transform based, such as discrete wavelet (DW) [7], Shearlet [8], discrete cosine (DC) [9], isotropic diffusion filtering [10], bilateral filters [11] are the most widely studied on conventional SAR image despeckling techniques. In these techniques, signals can be well estimated by a linear combination of few basis elements (i.e., the signals are sparsely approximated in transform domain), where scarce high-magnitude transform coefficients are retained and the rest are rejected due to certain noise. Block-matching and 3D (BM3D) [12] are the combination of transform domain and spatial domain image despeckling method, which are considered as a state-of-the-art in SAR image despeckling and are very well engineered methods. And, basing on sparse model and compressive sensing theory (CS), CS3D [13] despeckling framework is comprise of universal BM3D method. Also, Xu Bin [14] introduced sparse and redundant representation over learned dictionaries where the proposed algorithm at once train a dictionary on its corrupted content using K-SVD, however, it would also be very expensive, requiring the prolonged help of several expert interpreters under controlled conditions, and non-replicable, because of its inherently subjective nature.

Recently, the breakthrough of deep learning resulting from deep architecture has set off a wave in many fields (e. g., computer vision, pattern recognition [15-17]. Denoising autoencoder (DAE) [18] and convolutional denoising autoencoders (C-DAE) [19] are a recent addition to image denoising research. They are introduced by Vincent [18] as an extension of classical autoencoder and which are extensively studied for image denoising. The model attempted to learn the denoised image from its noisy version by some stacked layers, but they are not robust enough to variation in noise types beyond what it has seen during training. However, to the best of our knowledge, the above methods have a common formulation,

$$z = x + y \quad (1)$$

where speckle formulation is $lgz=lgx+lgv$, $z$ is the noisy image (observertation) generated as a combination of image $x$ and some noise $y$, all having image units. According to this formulation, most existing image denoising techniques attempt to approximate the clean image $x$ from a noisy observation $z$.

Deep neural networks can provide a competitive performance if the model is easy to access to very large scale datasets. Obviously, there are limited datasets are available to train the deep model. Therefore, it is still an open research area where applying deep learning application to image processing problems (for example denspeckling) with small dataset. Following reference [20], with our own modified architecture, this paper proposes SAR image despeckling model based on C-DAE with novel approach. We design more convolutional layers in the network architecture. Next, as shown in Fig. 1, the despeckling images are obtained by dividing the encoded map from the noisy observation. Finally, clean images are recovered using exponential transform. Through the feature learning process, batch normalization (BN) is incorporate into layers to improve the model accuracy and accelerate training speed from L2 to L7. Our proposed SAR image despeckling phase is illustrated in Fig. 1.

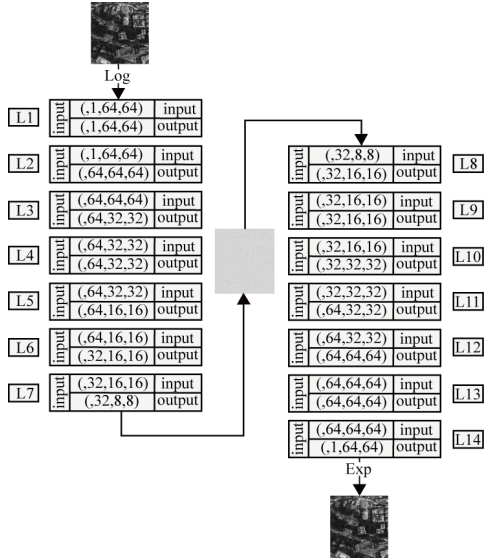

Fig.1 C-DAE phase of the model for SAR image despeckling.

Layers from L2 to L7 is encoding procedure while layers from L8 to L13 is decoding procedure. The contribution of our work in this paper can be summed up as follows:

1. We design a deep feed-forward back-propagation convolutional neural network model which split the noise from a noisy image, and batch normalization is also incorporated to boost model performance.

2. It is a novel model that despeckling the sar image with deep learning approach.

Rest of the paper is organized as follows. Next section discuss some preliminaries and our proposed model. Our experimental setting and details are introduced in Sect.3 and finally Sect. 4 elaborates our conclusions.

I. PRELIMINARY AND PROPOSED MODEL

In this section, we have provided preliminaries of the SAR image despeckling technique and the proposed method.

A. Autocoder and its variation

An autoencoder is a type of neural network that tries to learn an approximation to input using back-propagation (BP), given a set of training inputs $x_1,x_2,..., x_n$, it uses

$$z_i \approx x_i \quad (2)$$

An autoencoder first takes an image (input) $x \in [0,1]^d$ and encode it to a hidden representation $y \in [0,1]^d$ using a deterministic function,

$$y = s(Wx + b) = f_\theta(x) \quad (3)$$

where s is a nonlinear activation function (such as Sigmoid, ReLu), $W$ and $b$ are weight and bias, respectively. The latent (hidden) representation $y$ is then decoded into a reconstruction $z$, something is of same shape as original image $x$ by,

$$s(u) = 1/(1+e^{-u}) \quad (4)$$

$$z = s(W'y + b') = g_{\theta'}(y) \quad (5)$$

The parameters $\theta=\{W,b\}$, $\theta'=\{W',b'\}$ are optimized by minimizing an appropriate cost function. $\theta'$ is the transpose of $\theta$. [21] show a basic architecture of an autoencoder.

Denoising autoencoder is an extension of classical autoencoder which force the model to learn reconstruction of input given its noisy version. Formally, let $x_i$ $(i =1, 2, . . . , N)$ be the original image and $y_i$ $(i = 1, 2, . . . , N)$ be the corresponding noisy version of each $x_i$, then DAE can be presented as follow:

$$h(y_i) = s(Wx_i + b) \quad (6)$$

$$\hat{x}(y_i) = s(W'h(y_i) + b') \quad (7)$$

where $h_i$ is the hidden layer activation, $(y_i)$ an approximation of $y_i$ and $\{\theta,\theta'\}$ are still parameters to be optimized by appropriate cost function. Basic architecture of a donoising autoencoder is shown in [22].

Convolutional autoencoder is an extension of conventional autoencoder with convolutional encoding and decoding layers, differing from autoencoder as its weights are shared among all input location, which help to preserve spatial locality. Currently, medical image desnoising based on convolutional autoencoder had better performance than the traditional denoising autoencoders [23]. Therefore, we compared our results with SAR image despeckling based on traditional denoising autoencoder.

$$h_i = s(x \divideontimes W_i + b_i) \quad (8)$$

where ※ means 2D convolution. Reconstruction $y$ is given as

$$y=s(\Sigma h_i ※ W_i'+b_{ic}) \quad (9)$$

where $W_i'$ is the flip operation, $b_{ic}$ is bias of input channel $c$.

### B. Batch normalization

Batch normalization (BN) is considered as the excellence for convolutional neural network (CNN). We incorporate BN for SAR image C-DAE. Assume that $x \in R^d$ is the batch of images to the first layer of a networks with dimension d .Then, each dimension of x is normalized by

$$\hat{x}_i=(x_i-E(x_i)) / SQRT(VAR[x_i]) \quad (10)$$

where $E(x_i)$ is an expectation of xi and $VAR[x_i]]$ is the variance of $x_i$, and they are computed over the training period. SQRT is the square-root function. This type of normalization speedup convergence, even when the features are not decorrelated. Usually, the really represented layers may be changed via normalizing each input sometimes. To overcome the obstacle, the transformation inserted in the network is selected to identity it. More formally, a pair of parameters $\gamma_i$ and $\beta_i$ for each xi have been introduced to scale and shift the normalized value as:

$$y_i= \gamma_i \hat{x}_i+\beta_i \quad (11)$$

where γi and βi are learned along with the model parameters. In such manner or formulation, CNN model is benefited from batch normalization.Batch normalization helps the network train faster and provides higher accuracy. The result of Reference [23] shows the test accuracy of MNIST network after batch normalization is incorporated in the model. For our SAR image despeckling, we found that incorporating BN between convolution layer and ReLu activation function boost the accuracy of the model and make the training process faster where ReLu is an activation function defined as,

$$R(x) = max(0, x) \quad (12)$$

## II. EXPERIMENTAL SETTING

### A. Training and testing data

All the SAR image prepossessing techniques will be applied to Sentinel1 sensor image from the European Space Agency (ESA). Reasonably, all images are re-sized to 64×64 for computational resource. It is hard to get the ground truth of public SAR images. Simulation experiment is carried out by add various gamma noise (speckles). We train the model for despeckling with an interval of 0.1 for noise level $\sigma$, ranging from 0.1 to 1 independently, and the mean of noise $\mu$ is 0. Otherwise, it is not easy to compare our result with other methods. Randomly, 15 among 300 images, which are not trained, are chosen to test the despeckling quality.

Tensorflow[24] is used to set up the model with Relu and BN shown in Fig. 1 on an Inspur server (Intel Xeon ,64GB RAM, no GPU). Images are compared using peak singal to noise ratio (PNSR) and structural similarity index measure (SSIM) for coded images and output images [25, 26]. PNSR is a basic judgment that whether a despeckling method is providing a good performance or not. SSIM is used as the measurement or prediction of image quality which is based on an initial uncompressed or distortion-free image as reference Higher the PSNR means better image quality. It is the same with SSIM.

### B. Empirical evaluation

SAR images corrupted with certain noise level (σ=0.1) were used for baseline comparison. 300 SAR images are used to train model from dataset.with 12 testing sets left.

In the Fig. 2, the first column is the original images,from second column to fifth column is output images using BM3D, DAE, C-DAE,C-DAE with BN. Fig. 2 shows an increase despeckling result using a series of DAE network over the BM3D, which is the state-of-the-art conventional method. However, BM3D is based on block that smooth the image locally. TABLEI and TABLEII show the qualitative and quantitative comparisons respectively. Fig. 3 shows the loss function value linking to the epochs. It can be seen that even using half epochs, costing half time, both C-DAE methods would have achieved similar loss result with train loss when validating the samples. However, C-DAE with BN can get the final value much quicker when begin computing the train loss.

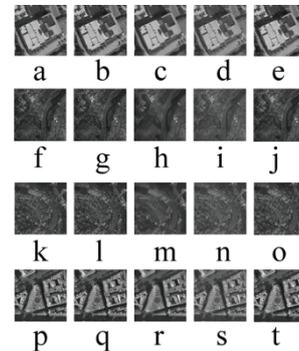

Fig. 2 Four samples are chosen to display the processing result

TABLE I. Mean PSNR and SSIM for full dataset (σ=0.1, samples=12)

| Image type | PSNR | SSIM |
|---|---|---|
| Noisy | 36.45 | 0.82 |
| BM3D | 36.32 | 0.88 |
| DAE | 39.01 | 0.92 |
| C-DAE | 40.83 | 0.94 |

TABLEII. Comparison for extremely high noised images (σ=1.0)

| Image | PSNT | SSIM |
|---|---|---|
| Noisy | 25.01 | 0.72 |
| BM3D | 21.82 | 0.78 |
| DAE | 27.01 | 0.82 |
| C-DAE | 28.83 | 0.84 |
| C-DAE with BN | 28.83 | 0.84 |

To test the ultimate performance of C-DAE with BN, 3 randomly chosen images with varying noise levels are used. Only building contour is seen from high corrupted images. It is of interest for despeckling performance. BM3D method falls into a local adaptation that in accordance with Fig. 2. The comparison of C-DAE with some other method for images corrupted with extremely high level noise are shown in TABLE II. However, regardless of time consuming, there is no different between the C-DAE and C-DAE with BN.

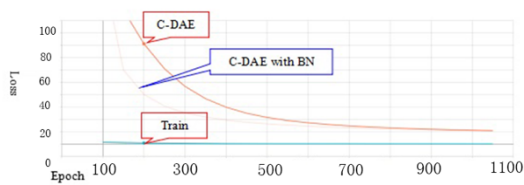

Fig. 3 Loss from 1000 epochs using a batchsize 10

## III. CONCLUSION AND FUTURE WORK

We have shown that efficient convolutional denoise autoencoder model with batch normalization for small train dataset is proposed. We test with denoise autoencoder and its variation on different images. They have been tested with various noise levels.

Comparing with transform base method, the proposed method is computing directly in image (spatial) domain. The model is a single process and performs better than others when evaluated by SSIM and PNSR. The proposed model is trying to learning the noisy images to estimate a clean image. This SAR image despeckling process basing deep networks also provide experience for other SAR image understanding.

ACKNOWLEDGMENT

This research was funded in part by National Key R&D Program of China, grant number 2016YFB0501805.